\documentstyle[psfig]{mn}
\newcommand{\caii}{Ca~{\sc ii}}
\newcommand{\cafii}{[Ca~{\sc ii}]}
\newcommand{\feii}{Fe~{\sc ii}}
\newcommand{\fei}{Fe~{\sc i}}
\newcommand{\Oi}{O~{\sc i}}

\newcommand{\oi}{[O~{\sc i}]}
\newcommand{\oii}{[O~{\sc ii}]}

\newcommand{\ha}{H$\alpha$}
\newcommand{\hb}{H$\beta$}
\def\lsim{\!\!\!\phantom{\le}\smash{\buildrel{}\over
 {\lower2.5dd\hbox{$\buildrel{\lower2dd\hbox{$\displaystyle<$}}\over
                              \sim$}}}\,\,}
\def\gsim{\!\!\!\phantom{\ge}\smash{\buildrel{}\over
 {\lower2.5dd\hbox{$\buildrel{\lower2dd\hbox{$\displaystyle>$}}\over
                              \sim$}}}\,\,}

\title[Circumstellar interaction of SN 2002ic]
{Circumstellar interaction of the type Ia supernova 2002ic}

\author[N.N. Chugai, R.A. Chevalier, P. Lundqvist]{N.N. Chugai$^1$,
 R.A. Chevalier$^2$, P. Lundqvist$^3$\\
$^1$Institute of Astronomy, RAS, Pyatnitskaya 48, 109017 Moscow,
 Russia\\
$^2$Department of Astronomy, University of Virginia, P.O. Box 3818, Charlottesville, 
 VA 22903, USA\\
$^3$Stockholm Observatory, AlbaNova, Department of Astronomy, SE-106~91 Stockholm, Sweden
}

\begin{document}

\maketitle

\begin{abstract}

We propose a model
to account for the bolometric light curve, quasi-continuum 
 and the \caii\ emission features  of the peculiar type Ia supernova (SN) 
2002ic, which exploded in a dense circumstellar envelope. 
The model suggests that the SN~Ia 
had the maximum possible kinetic energy and that the ejecta expand
in an approximately
spherically symmetric (possibly clumpy) circumstellar environment.
The \caii\ and quasi-continuum are emitted by shocked SN ejecta 
that underwent deformation and fragmentation in the intershock layer. 
Modeling of the \caii\  triplet implies that the 
contribution of the \Oi\ 8446 \AA\ line is about 25\%  of
the 8500 \AA\ feature on day 234, which permits us to 
recover the flux in the \caii\ 8579 \AA\ triplet from the
flux of 8500 \AA\ blend reported by Deng et al. (2004).
We use the \caii\ doublet and triplet fluxes on day 234
to derive the electron temperature  ($\approx4400$~K) 
in the \caii\ line-emitting zone and the ratio of the total
area of dense fragments to the area of the shell, $S/S_0\sim 10^2$.
We argue that 
\caii\ bands and quasi-continuum originate from  
different zones of the shocked ejecta that reflect the 
abundance stratification of the supernova. 

\end{abstract}


\section{Introduction} \label{sec-intro}

Recently,   strong evidence has been presented 
that a  sub-class of type IIn supernovae (SN) (`n' stands 
for narrow \ha) includes rare SN~Ia events   exploding in a 
dense circumstellar medium (CSM) (Hamuy et al. 2003). To date, the new variety
comprises three events: 
SN~2002ic, SN 1997cy and SN 1999E (cf. Deng et al. 2004).
Statistical arguments indicate that the new family constitutes 
less than one percent of all SN~Ia (Chugai \& Yungelson 2004).
All these supernovae remain very luminous after the light curve maximum,
which is related to the CS interaction.
A light curve model of SN~1997cy suggests that 
the CS envelope within $3\times10^{16}$ cm contains several
solar masses (Chugai \& Yungelson 2004). 

Two 
options for the origin of the CSM have been proposed (Hamuy et al. 2003): 
mass-loss by a supergiant in a binary scenario of SN~Ia 
(Whelan \& Iben 1973) or the mass-loss by a single supergiant
in the SN\,1.5 scenario (Iben \& Renzini 1983).
The similarity of SN~2002ic events seems to favour the SN\,1.5 scenario, 
with the progenitor mass around $8~M_{\odot}$ 
(Chugai \& Yungelson 2004). The possible scenario of a 
C/O white dwarf (WD) merging with the CO-core of a supergiant companion 
(Livio \& Riess 2003) faces a serious problem 
in explaining the short time lag between the ejection of the 
common envelope and the explosion (Chugai \& Yungelson 2004). 

The study of the CS interaction and modeling of the spectra 
should elucidate the late evolution and the 
origin of these interesting events.
An important recent conclusion is
that SN~2002ic interacts with an equatorially 
concentrated CSM (Deng et al. 2004). This statement
is based upon the interpretation that line profiles 
in SN~2002ic are too broad compared with the predictions of a
spherical model. 
Also, spectropolarimetric observations of SN~2002ic reveal the presence 
of polarization which is likely to be related to the CSM 
(Wang et al. 2004). The claimed asymmetry of the CSM may have 
an important implication for the problem of 
the evolutionary scenario, because the single and binary scenarios 
are likely to possess different degrees of symmetry. 

The aspherical model predicts variations of the
emission line profiles of SN~2002ic-like 
supernovae viewed from different angles. However, all three
known objects (see Deng et al. 2004, their Fig. 1)  show 
similar shapes and widths of spectral features.
This fact implies that, despite the detected polarization 
of the emission from SN~2002ic, 
the interaction possesses a high degree of sphericity.
In view of this, we investigate a spherical model here.
We attempt to explain the light curve, the \caii\ line profiles and the 
quasi-continuum of SN~2002ic-like events on the basis of 
a circumstellar interaction model. We rely 
on the spherical approximation, although we allow for local deviations 
from sphericity (e.g. clumpy structure).
We start with the simulation of the CS interaction dynamics and the bolometric 
light curve of SN~2002ic using the thin shell approximation (Section 2).
Although the model is the same as before 
(Chugai \& Yungelson 2004), we include more realistic ejecta abundances, 
which affects the light curve. 
We then analyse the line formation in the inhomogeneous 
intershock layer and propose 
a simple treatment of this problem (Section 3). Using 
this approximation we simulate the \caii\ line profiles and the quasi-continuum
of SN~2002ic with the kinematic parameters of the interaction model.
We demonstrate that the spherical model successfully 
accounts for the basic observational properties of SN~2002ic 
and discuss implications.

\section{CS interaction and the light curve} \label{sec-inter}

The interaction of a SN~Ia with a dense spherical
CS envelope and the light curve formation is calculated 
using the model applied earlier for the analysis of the 
CS interaction of SN~2002ic (Chugai \& Yungelson 2004). 
We recapitulate the basic assumptions of the model and emphasize 
some modifications.

\subsection{Model}\label{sec-model}

The density distribution in a freely expanding SN~Ia ejecta 
($v\propto r$) is approximated by the exponential law
$\rho\propto \exp\,(-v/v_0)$ with $v_0$ determined by the 
the mass $M=1.4~M_{\odot}$ and kinetic energy $E$. 
An exponential law better represents the density 
distribution in SN~Ia ejecta than a power law (Dwarkadas \& Chevalier 1998).
In an attempt to provide the maximum expansion velocity of the 
interaction shell we consider a SN~Ia  model with the maximum
kinetic energy, corresponding to the detonation model DET1 of 
Khokhlov et al. (1993):
$E=1.75\times10^{51}$ erg with the $^{56}$Ni mass 
of $M_{\rm Ni}=0.92~M_{\odot}$.  
We do not claim that the DET1 detonation regime  actually occurs 
in SN~2002ic. This model strongly underproduces 
intermediate mass elements (e.g. Ca) 
(Khokhlov et al. 1993), which is at odds 
with the strong \caii\ lines observed in SN~2002ic (see below Sec. 3.4). 
Parameters of the DET1 model are adopted  because this 
case illustrates what happens if all  of a Chandrasekhar mass 
C/O WD is  burned. 
The important point here is that the complete incineration of 
C and O is not required: if the burning in the outer layers is terminated at 
the intermediate elements, the released nuclear energy is practically
 the same as in the case of complete incineration.
 
The CS density distribution is described by  
power laws ($\rho\propto r^{k}$) with  different power law indices
for each of three suggested radial zones: $k_1$ in the range $r<r_1$,
$k_2$ for $r_1<r<r_2$, and $k_3$ for $r>r_2$. The variation 
of the power law index is suggested by two observations: (i)
the luminosity decline after about day 400 (Deng et al. 2004)
suggests a CS density drop beyond
some radius $r>r_2$; (ii) the rising early 
visual luminosity of SN~2002ic (Wood-Vasey et al. 2002;
Hamuy et al. 2003) suggests 
a flat (or rising outwards) CS density distribution in  the inner part
($r<r_1$) of the CS envelope (Chugai \& Yungelson 2004).

The hydrodynamic interaction is calculated in the thin 
shell approximation, in which the layer between 
the forward and reverse shocks is replaced by a geometrically 
thin shell (Chevalier 1982b). 
The kinetic luminosities of the forward and
reverse shock waves are converted into X-ray luminosities 
using transformation coefficients dependent 
on cooling rates of the postshock gas and the expansion time (Chugai 1992;
Chevalier \& Fransson 1994). For the  CS density range of interest,
the reverse shock is always radiative, which leads to 
the formation of a cool dense shell (CDS) composed of the shocked 
SN ejecta.
The forward shock 
can be also radiative and contribute to the CDS formation.
The X-ray radiation of both shocks absorbed by the 
cool material (SN ejecta, CDS, and CSM) is 
identified with the optical bolometric luminosity powered 
by the CS interaction. 
The additional luminosity powered by the radioactive decay
$^{56}$Ni -- $^{56}$Co -- $^{56}$Fe
is calculated using an analytical theory for the SN light curve 
(Arnett 1980).
The resulting light curve is the  sum of the interaction and 
radioactive luminosities.

 \begin{figure}
  \centering
  \vspace{9 cm}
  \includegraphics{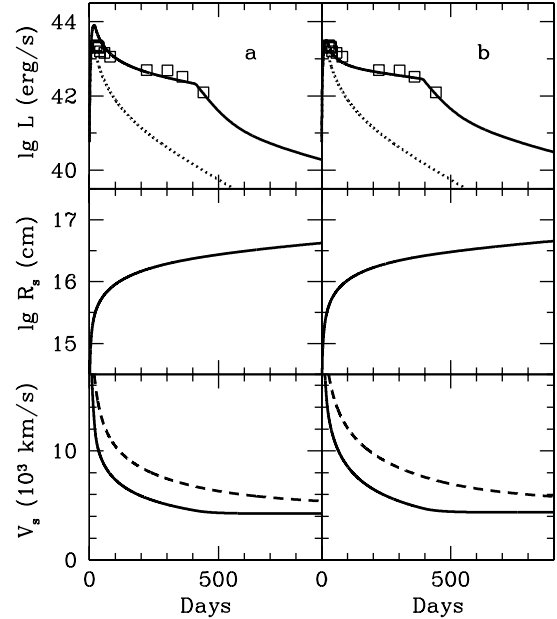}
  \caption[]{Bolometric light curve (upper panels), radius 
  of the thin shell (middle), and expansion velocity (bottom) 
  for models with CS density $\rho\propto r^{-2}$ ({\em a}) 
  and $\rho\propto r^{-1.4}$ ({\em b}). The bolometric 
  light curve with CS interaction ({\em solid} line) and without 
  CS interaction ({\em dotted}) are overplotted on the data of Deng et al. (2004).
  The lower panel shows the velocity of the thin shell ({\em solid}) and 
  the SN boundary velocity ({\em dashed}).
       }
  \label{f-blc}
  \end{figure}

An important modification in the present version is 
the inclusion of a `realistic' chemical 
composition of the SN~Ia ejecta instead of the previously assumed
solar composition (Chugai \& Yungelson 2004). 
The SN~Ia composition is now approximated  by
a mixture of Fe and Si, where the amount of Fe 
is equal to the ejected $^{56}$Ni mass. The new composition results in 
more efficient cooling in the reverse shock and more 
efficient absorption of the X-rays in the SN ejecta.
The adopted absorption  coefficient for X-rays in the SN material 
is $k_X\approx500E_{\rm keV}^{8/3}$ cm$^2$ g$^{-1}$, i.e., 50 
times larger than that with a solar composition. 
Owing to the efficient absorption of X-rays in the SN~Ia ejecta with 
the `realistic' composition, the rate of emission of 
optical photons per unit of kinetic luminosity is higher in 
the present model compared with the earlier version. 

\subsection{Light curve}\label{sec-lc}

\begin{figure}
  \centering
  \vspace{6 cm}
  \includegraphics{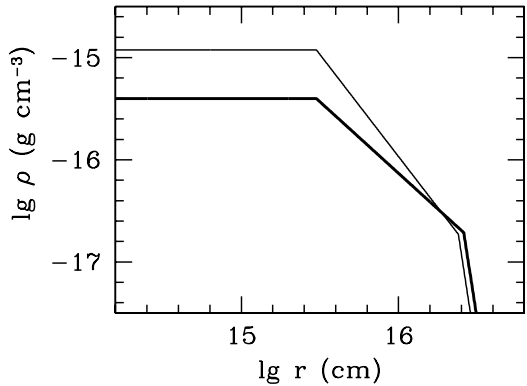}
  \caption[]{Circumstellar density in the model with 
 $\rho\propto r^{-2}$ ({\em thin} line)  and $\rho\propto r^{-1.4}$ 
 ({\em thick} line). In both models, the inner zone has a plateau 
 density distribution to provide the rise of the light 
 curve to the maximum at about day 20.
       }
  \label{f-rho}
  \end{figure}

Our modeling of the bolometric optical light curve 
is compared with the SN~2002ic data from Deng et al. (2004),
modified here for a Hubble constant of
$H_0=70$ km s$^{-1}$ Mpc$^{-1}$.
Two versions of the light curve computations along with the 
corresponding shell radius and velocity 
are presented in Fig.~\ref{f-blc}. 
Fig.~\ref{f-blc}a shows the model with
$k_2=-2$, whereas the second model (panel {\em b})
corresponds to $k_2=-1.4$.
Both CS density distributions are shown in Fig.~\ref{f-rho}.
The model with $k_2=-1.4$  provides the largest expansion 
velocities of the thin shell ($v_{\rm s}$) and of the SN at the 
boundary ($v_{\rm sn}$).   
 From the point of view of the interpretation of the
 observed broad line profiles,
this model is preferred. Henceforth, this case is 
referred to as the standard model. 

The flatter CS density distribution of the model 
compared with a steady-state 
wind implies that after the main mass loss episode,
the mass loss rate gradually decreased while the
presupernova evolved towards the explosion.
The total mass of the CS envelope in the standard model is 
$1.6~M_{\odot}$. This estimate should be 
considered  a lower limit, because we neglect the possible wind 
at $r\gg r_2$ that  may comprise a significant mass,
despite its lower density. 
If the wind velocity out to $r_2\approx 2.6\times 10^{16}$ cm is
$> 10$ km s$^{-1}$, then the age of the strong mass loss is $<820$ yr
and the mass loss rate is $>2\times 10^{-3}~M_{\odot}$ yr$^{-1}$.

We recalculated the light curve of SN~1997cy for the detonation model 
with a `realistic' composition and found that the mass of the CS envelope 
in this case is $\approx2.5~M_{\odot}$, a factor two lower than our 
previous estimate for SN~1997cy based upon a model with lower 
explosion energy and normal composition of the SN ejecta 
(Chugai \& Yungelson 2004). Most of the difference is related to the 
efficient conversion of the kinetic luminosity into the bolometric luminosity
owing to the efficient absorption of X-rays by the ejecta.

\begin{figure}
  \centering
  \vspace{7.5 cm}
  \includegraphics{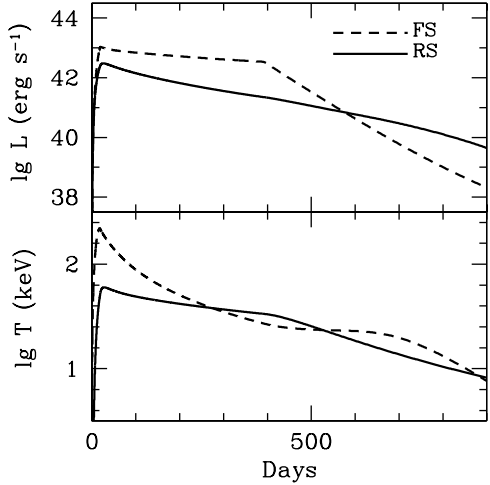}
  \caption[]{Luminosity of escaping X-rays (top)
  and temperature (bottom)
  of the reverse shock ({\em solid} line) and forward shock ({\em dashed})
       }
  \label{f-xlc}
  \end{figure}

The model of the optical light curve predicts the  
luminosity and flux of the escaping 
X-rays, which may be used as an additional test of the model for 
SN~2002ic-like events. 
In Fig. \ref{f-xlc}, we plot the total luminosity of the escaping X-rays
 from the forward and reverse shocks, together with the
 temperatures of both shocks for the standard model. 
Due to the high CS density and rather flat CS density distribution,
the velocity jump at the inner shock is relatively high (Fig.\ref{f-blc}). 
This, combined with 
the higher molecular weight of the SN~Ia matter, results in a high
temperature of the reverse shock that is comparable to the 
temperature of the forward shock after day 250. In a SN~II 
interacting with a wind, the reverse shock is usually
 markedly slower and cooler
than the forward shock (Chevalier 1982b). 
Remarkably, the luminosity of the forward shock is large and 
provides a substantial contribution to the optical SN luminosity.
This is due to the fact that the forward shock is 
radiative: it enters the fully radiative regime on 
day $\sim150$. The reverse shock is  radiative through the 
time period we have modeled. The inflection in the 
forward shock luminosity on day $\sim 400$ 
reflects the corresponding inflection in the 
adopted  CS density distribution at $r\approx 2.6\times10^{16}$ cm 
(Fig. \ref{f-rho}).
The forward shock becomes nonradiative soon after that point, at 
$t>440$ d.

\begin{figure}
  \centering
  \vspace{6.5 cm}
  \includegraphics{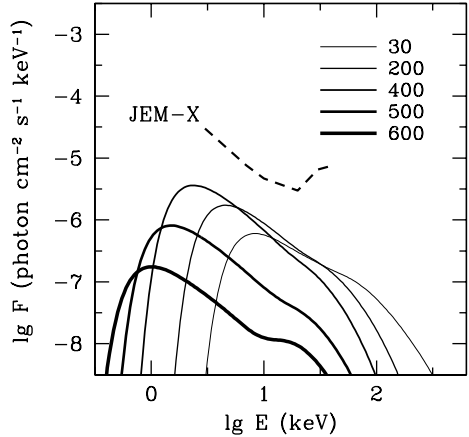}
  \caption[]{X-ray photon flux from SN~2002ic predicted by the
  standard model
  at different epochs between days 30 and 600 assuming $H_0=70$
  km s$^{-1}$ Mpc$^{-1}$ ($D=285$ Mpc). The $3\sigma$
  sensitivity
  curve of JEM-X monitor of INTEGRAL ({\em dashed} line) assumes a $10^6$ s
  integration time.
       }
  \label{f-xfc}
  \end{figure}

The photon spectrum of the emergent X-ray flux at different epochs 
between day 30 and day 600 is shown in Fig. \ref{f-xfc} together with the 
detection limit of the
JEM-X monitor of the INTEGRAL satellite assuming a
3$\sigma$ detection level with the integration time of
 $10^6$ s (http://astro.estec.esa.nl/Integral/).
The plot demonstrates that the  detection of X-rays from 
SN~2002ic is not expected in
our model for a reasonable integration time. On the other hand, 
a SN~2002ic-like event exploding at a distance 
$D\leq100$ Mpc would be detectable at the favorable epoch of 200--400 days
by the JEM-X monitor with an integration time $\leq10^6$ s.
Chandra and XMM may be even more efficient in detecting the X-rays 
from  distant SN~2002ic-like events. 
Between days $200-500$ the predicted flux of SN~2002ic in 
the $0.5-10$ keV band is $\sim 10^{-14}$ erg s$^{-1}$ cm$^{-2}$.
Given the Chandra detection limit 
$\sim 10^{-15}$ erg s$^{-1}$ cm$^{-2}$ in the $0.5-10$ keV band, 
we  conclude that X-rays from SN~2002ic could have been detected 
at an age $200-500$ days. At  later epochs,
the flux quickly decreases (Fig. \ref{f-xfc})
and on day 600 it becomes far below the Chandra sensitivity. 

The above model for the escaping X-ray flux 
assumes a smooth CSM. Clumpiness may modify these results
in two ways: a different temperature in the forward shock, and 
a different escape probability for photons through the clumpy CSM.

\section{Formation of the optical spectrum}\label{sec-linef}

The late time Subaru spectrum  on day 
$234$  (assuming the explosion date JD 2452585)
consists of broad emission bumps (we dub them quasi-continuum) with  
superimposed emission lines of \ha, \hb, \caii\ 3900 \AA,
and  \caii\ 8600 \AA\ blended with \Oi\ 8446 \AA\ (Deng et al. 2004). 
Most of the bolometric luminosity in our dynamical model comes 
from the shocked ejecta. We, therefore, believe that, apart from 
\ha\ and \hb\ and other narrow lines, 
most of the optical spectrum on day 234 
(i.e. quasi-continuum and \caii\ lines) originates 
from shocked SN ejecta. This conjecture will be checked below.

We only briefly address the important issue of \ha\ line 
formation. 
Although the CS origin of the narrow \ha\ emission is certain,
the nature of the broad \ha\ is unclear.
Multiple Thomson scattering as a mechanism for the broad \ha,
applied earlier to SN~1998S (Chugai 2001), and 
discussed by Wang et al. (2004) in connection with SN~2002ic, 
requires a large optical depth $\tau_{\rm T}\sim 3$.  This is 
in conflict with our dynamical model that predicts 
a Thomson optical depth of the undisturbed  
CS ejecta of $\tau_{\rm T}\approx 0.1$ on day
$\approx 240$ .
In our opinion, the broad \ha\ comes from the shocked cool dense 
hydrogen. There are three possible origins for that line-emitting gas: 
(i) the outer layers of the SN ejecta may contain the remains of the 
hydrogen envelope of the presupernova (in the SN~1.5 scenario), 
so the cool shocked ejecta enriched by hydrogen 
may be partially responsible for the 
broad \ha\ component; (ii) the forward shock becomes radiative 
in the case of smooth CS hydrogen after about day 150, so 
the shocked CS hydrogen cools at that epoch and becomes able to 
emit broad \ha; (iii) in the case of a clumpy CS envelope 
radiative shocks in CS clouds and the shear flow around the shocked clouds
may produce the broad velocity spectrum of the line-emitting gas that 
can be responsible for the intermediate and broad \ha\ line.
In reality, all three  sources of the cool shocked 
hydrogen may contribute to the broad \ha\ component.
However, only in the third option can the observed intermediate \ha\
component with a width of $\sim2\times10^3$ km s$^{-1}$
(Hamuy et al. 2003)  be produced. 

Henceforth, we concentrate on the study 
of the \caii\ emission lines and the quasi-continuum.

\subsection{Line formation model}\label{sec-mech}

Momentum and mass conservation
imply a very small thickness of the CDS,
of the order of $\delta\sim R(v_{\rm sh}/c_{\rm s})^2$, where
$v_{\rm sh}=v_{\rm sn}-v_{\rm s}$ is the velocity of the reverse
shock and $c_{\rm s}$ is the ion sound speed in the cool gas.
With $v_{\rm sh}\approx 3000$ km s$^{-1}$
at an age of 230-250 d (see Fig. 1), adopting
the temperature of the cool gas of $10^4$~K and the ion average atomic
weight $\approx 40$
(i.e.,  $c_{\rm s}\approx2$ km s$^{-1}$), $\delta /R \sim 10^{-6}$.
For such a thin shell the velocity dispersion along the line of sight
is negligible, so the line radiation transfer in this shell
may be treated in the static approximation (Chevalier and Fransson 1994).
The expansion effect appears only in the transformation
of the shell emission to the observer frame.
However, the picture of a spherical thin CDS
is an idealization;
in reality, as we will show below, the shell should be strongly
distorted.

\subsubsection{Line-emitting shell}\label{sec-shell}

A straightforward argument against the existence of a
geometrically thin spherical shell is based
upon the observed profiles of the \caii\ features.
It is well known that in the optically thin case the line
profile from a geometrically
thin expanding shell should be boxy. This might be a
sensible approximation
for the observed \caii\ features given some unavoidable
perturbation and smearing of the boxy profile.
The problem, however, is that the \caii\ lines are expected to be optically
thick, not thin.

To show that the dense shell is  opaque in the \caii\ lines,
we consider the conditions
on day 240.  The swept up SN mass in the reverse shock in the standard
model is $\approx0.6~M_{\odot}$ and the shell radius is
$R\approx 1.8\times10^{16}$ cm. Assuming that the swept up mass
resides in the CDS and that the Ca abundance among Fe-peak elements
is solar, i.e.,  $\approx0.03$ by mass (a sensible approximation for 
a rough estimate) one gets the column density of \caii\ in
the CDS  of $N(\mbox{CaII})\sim 2\times 10^{20}\phi_2$ (where $\phi_2$
is the ionization fraction of \caii). The optical depth in the \caii\
doublet lines is then
\begin{equation}
\tau=\frac{\pi e^2}{mc}\frac{f_{12}}{u_{\rm t}}N(\mbox{CaII})
\sim 10^9\phi_2\,,
\label{eq-tauca}
\end{equation}
where $f_{12}=1$ is the oscillator
strength for the \caii\ doublet and $u_{\rm t}\sim 2$ km s$^{-1}$
is the thermal velocity assuming $T=10^4$~K. On day 234,
the number density in the CDS
under pressure equilibrium
is $\sim 10^{13}$ cm$^{-3}$. Assuming LTE in the
cool dense gas we obtain $10^{-3} <\phi_2 < 1$
for $4000<T<10^4$ K. The equation (\ref{eq-tauca})
shows that the optical depth of the CDS in the \caii\ doublet is
very large for the estimated \caii\ ionization fraction.
It is easy to check that, given the low excitation potential of the
\caii\ term $^2$D, the \caii\ triplet should be optically thick
too.

Optically thick \caii\ lines from a geometrically
thin shell are expected to be double-horned (M-shaped)
$F\propto |\lambda-\lambda_0|$
with peaks at the shell velocity ($\pm v_{\rm s}$) (Gerasimovi\u{c} 1933).
This type of profile is apparently inconsistent with the
smooth, round-topped (flat-topped) \caii\ doublet
and triplet profiles observed in SN~2002ic (see Deng et al. 2004).
To resolve this controversy, we follow
Cid Fernandes \& Terlevich (1994) in suggesting  that the
round-topped  \ha\ line profile observed in SNe~II interacting
with a dense CS environment are produced by a thin shell corrugated
due to the Rayleigh-Taylor (RT) instability.

It is well known that the decelerated thin shell in the intershock layer
is subject to the
Rayleigh-Taylor (RT) instability (Chevalier 1982a; Nadyozhin 1985).
RT spikes grow and degrade via shearing and advective flows,
thus producing a two-fluid mixture consisting of dense cool
gas and hot postshock gas behind the forward shock wave.
As a result, a mixing layer with an extent $\Delta R/R\sim 0.1$ forms
between the thin shell and the forward shock as demonstrated by
2D and 3D simulations (Chevalier \& Blondin 1995; Blondin \& Ellison 2001).
The clumpy  structure of the CSM may cause additional perturbations
of the mixing layer and lead to the corrugation of the
forward and reverse shocks (Blondin 2001, his Fig. 6).
If both reverse and forward shocks are radiative, the shell
instability becomes even more vigorous due to the 
nonlinear thin shell instability (Blondin \& Marks 1996; Hueckstaedt 2003).
The above arguments based upon the profile shape and the interaction
hydrodynamics suggest that a realistic model of the intershock
line-emitting layer should include the
inhomogeneous distribution of the dense shocked gas.

\begin{figure}
  \centering
  \vspace{5 cm}
  \includegraphics{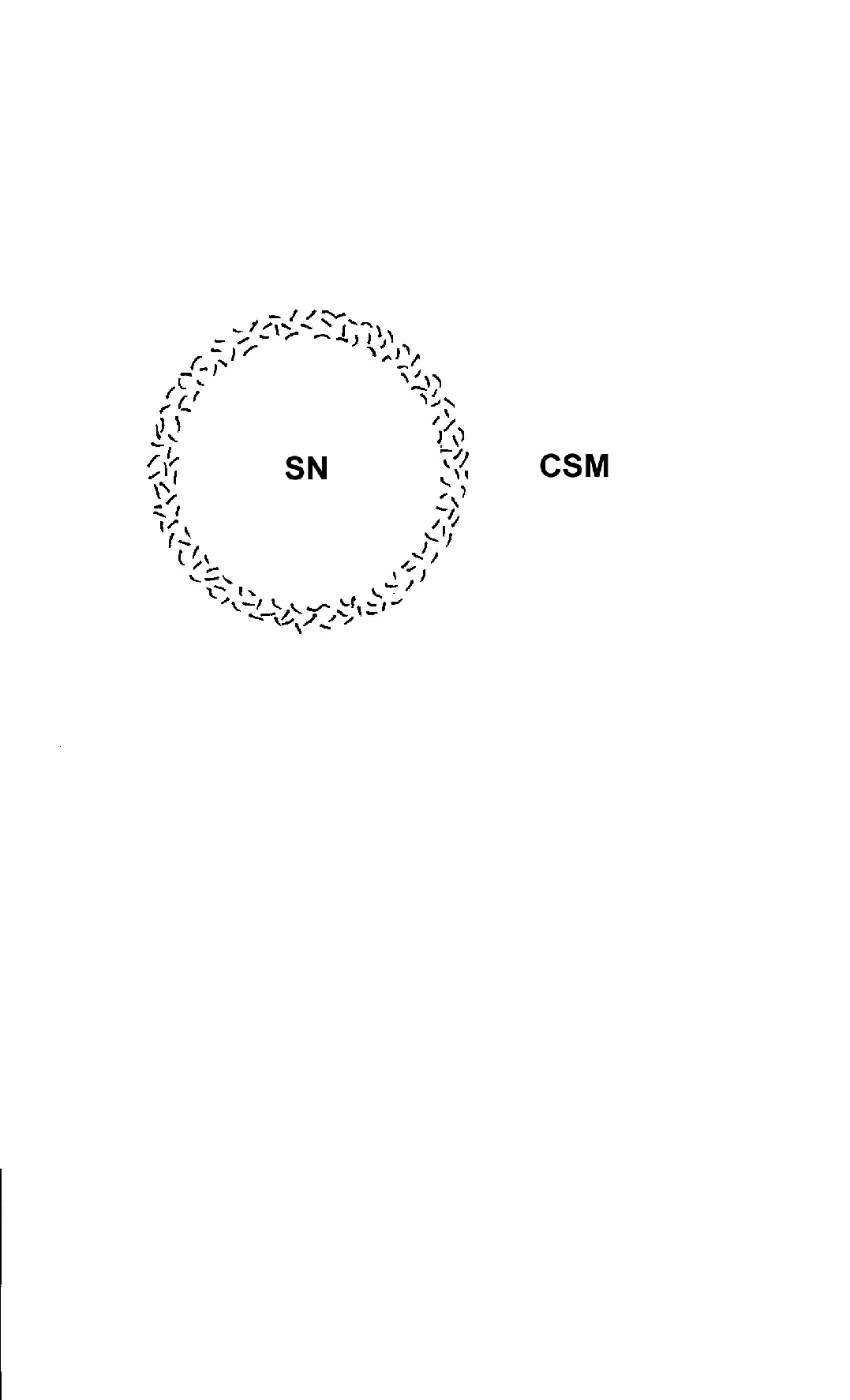}
  \caption[]{A schematic view of SN~2002ic that
  shows the fragmented cool dense shell responsible for the
  bulk of the optical radiation. The fragmented shell is
  bounded by the forward shock propagating in the CSM
  and by the reverse shock propagating in the unshocked SN ejecta.
              }
  \label{f-cart}
  \end{figure}

To treat the line formation in the
inhomogeneous intershock layer, we reduce it to
a spherical `shell' of a thickness $\Delta R=2\xi R=0.1R$
in the range of radii $|r-R|<\xi R$ filled with similar, randomly
oriented, dense planar fragments
(Fig. \ref{f-cart}). Each fragment is a thin sheet with an area
$\sigma$ and total surface area $2\sigma$.
As a result of the deformation and stretching of dense fragments,
their cumulative area $S$ should, generally, exceed the area
of the unperturbed
spherical dense shell $S_0=4\pi R^2$ by a factor $S/S_0> 1$.
As we will see below, this parameter (called the `area ratio')
turns out to be crucial for the interpretation of the high
luminosity of the \caii\ features in SN~2002ic. Moreover, 
the high
luminosity of the \caii\ features requires $S/S_0\gg 1$. This is the
reason why we omit in our simple model (Fig. \ref{f-cart})
the thin spherical shell that is produced and destroyed
continuously in the reverse shock.

\subsubsection{Line radiation transfer in the shell}\label{sec-transfer}

We consider two plausible velocity distributions of filaments
in the shell $R(1-\xi)<r<R(1+\xi)$: a constant velocity and
a linear law $v=v_{\rm s}(r/R)$.
Line photons emitted by randomly distributed dense fragments may escape
or be absorbed by another fragment of the shell. To describe
the line radiative transfer inside the inhomogeneous shell we apply the
local escape probability language adapted for the `gas'
of opaque macroscopic chunks of dense
material. Definitions of the local optical depth and emissivity
must be modified accordingly.

The total number of fragments in the shell is
$N=4\pi R^2(S/S_0)/\sigma$ and the
fragment number density is $n_{\rm f}=N/V=(S/S_0)/(\sigma\Delta R)$,
where $V=4\pi R^2\Delta R$.
The average cross-section of a fragment is
$\langle\sigma\rangle$ with the
ratio $\langle\sigma\rangle/\sigma=0.5$ for plane sheets and $\sim 0.5$
for deformed sheets. For instance, for a cylindrical surface
with a radius $a$ and a height
$h\ll a$ this ratio is $\approx0.5$ with the accuracy of $O(h/a)$.
The radiation escapes from an optically thick fragment in the
frequency band $2x\Delta \nu_{\rm D}$ assuming
a gaussian absorption coefficient
$k_{\nu}\propto \exp\,[-(\Delta \nu/\Delta \nu_{\rm D})^2]$, where
$\Delta \nu_{\rm D}=u_{\rm t}/\lambda$ and
$x$ is defined by the condition $\tau(x)=1$, i.e., $x=(\ln\,\tau)^{0.5}$.
A fragment absorbs the incident radiation in
the same frequency band, $2x\Delta \nu_{\rm D}$.

The analog of the Sobolev optical depth in our case is the local
occultation optical depth,
$Q=2xu_{\rm t}\langle\sigma\rangle n_{\rm f}|dv/ds|^{-1}$.
To take into account the finite optical depth of a fragment this expression
should be multiplied by the absorption probability
of a fragment $(1-\mbox{e}^{-\tau})$.
For a shell with constant velocity, the
velocity gradient along some direction is $dv/ds=(v_{\rm s}/R)\sin^2\theta$,
where $\theta$ is the angle between the photon wave vector and the radius.
The expression for $Q$ then reads 
\begin{eqnarray}
Q=\frac{x}{\sin^2\theta}\left(\frac{S}{S_0}\right)\left(\frac{R}
{\Delta R}\right)
\left(\frac{u_{\rm t}}{v_{\rm s}}\right)(1-\mbox{e}^{-\tau})\nonumber \\
=Q_0(1-\mbox{e}^{-\tau})/\sin^2\theta\,,
\label{eq-tauoc}
\end{eqnarray}
where $\langle\sigma\rangle/\sigma=0.5$ is used. For the linear velocity
case, $Q=Q_0(1-\mbox{e}^{-\tau})$. To estimate $Q_0$ let us adopt
$v_{\rm s}=6000$ km s$^{-1}$, $u_{\rm t}=2$ km s$^{-1}$ and $x\approx 3$.
Equation (\ref{eq-tauoc}) gives $Q_0\sim10^{-2}(S/S_0)$.
Therefore, the shell becomes optically thick ($Q_0 >1$) for an
area ratio $S/S_0>10^2$ provided $\tau\gg 1$.

\begin{figure}
  \centering
  \vspace{7.3 cm}
  \includegraphics{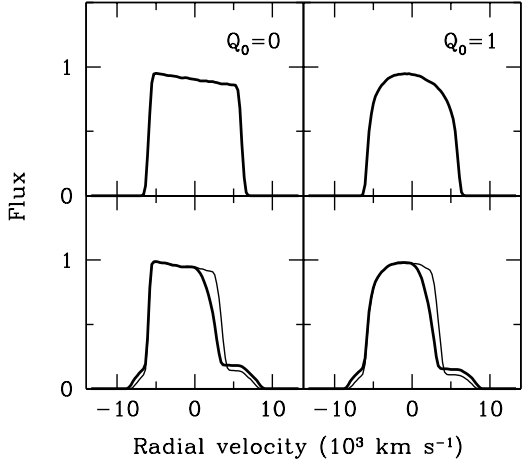}
  \caption[]{The line profile for two different models of the
  cold, dense shell. Upper panels show profiles
  for a transparent ($Q_0=0$) and a modestly opaque shell ($Q_0=1$).
  Lower panels show profiles for the same models, but
    with the addition of non-local scattering
  in the SN ejecta for a cut-off radius $r_{\rm c}=0.95$
  ({\em thick} line) and $r_{\rm c}=0.9$ ({\em thin} line).
       }
  \label{f-single}
  \end{figure}

The emissivity of the fragment ensemble is
$j=(1/2\pi)\sigma n_{\rm f}F$, where
$F$ is the frequency integrated flux. In the isothermal approximation
and assuming Boltzman populations, the flux is
$F=2\pi B_{\nu}(T)x\Delta \nu_{\rm D}(1-\rm{e}^{-\tau})$, so the emissivity is
\begin{equation}
j=\sigma n_{\rm f}B_{\nu}(T)x\Delta \nu_{\rm D}(1-\rm{e}^{-\tau})\,.
\label{eq-emiss}
\end{equation}

To compute the luminosity, one needs to take into account
self-absorption via the angle-averaged escape
probability $\beta=\langle[1-\exp\,(-Q)]/Q\rangle$.
Assuming pure absorption one gets the
luminosity $L=4\pi jV\beta$, where
$V$ is the shell volume $4\pi R^2\Delta R$.
With the line emissivity $j$ from equation (\ref{eq-emiss}), the
luminosity of \caii\ triplet is
\begin{equation}
L=16\pi^2 R^2(S/S_0) B_{\nu}(T)(1-{\rm{e}}^{-\tau})
\beta x\Delta \nu_{\rm D}\,.
\label{eq-lumir}
\end{equation}
The luminosity increases linearly with the area ratio for
$Q_0\ll 1$ (i.e., $\beta\approx 1$)
and saturates for large $S/S_0$, when $Q_0$ becomes
also large. Actually, for large $Q_0$ one gets
$\beta \sim 1/Q_0\propto S_0/S$ so $L$ becomes independent of $S/S_0$.
Remarkably, the line luminosity according to equation (\ref{eq-lumir})
exceeds the upper limit on the luminosity of a static shell by
a factor  $(S/S_0)\beta$ that, for large area ratio, may be
substantially greater than unity. This apparent paradox is related to the
velocity gradient that permits the volume emission regime for resonance
photons as opposed to the surface emission regime in the static case.
A finite albedo $\omega$ of a fragment for incident
photons requires a modification of the escape probability:
$\beta$ in equation (\ref{eq-lumir}) should be replaced by
$p=\beta/[\beta+(1-\beta)(1-\omega)]$.
However, we compute the line
profiles on the assumption of zero albedo, which, as we checked,
does not affect the profile significantly.

In the limit of a small area ratio ($S/S_0\sim 1$)
the luminosity of the \caii\ triplet on day 234 for the
probable temperature range $T<10^4$~K for $\beta=1$ is
$L(8579\mbox{\AA})<5\times10^{40}$ erg s$^{-1}$. On the other hand,
using the flux in the 8500 \AA\ band,
$f\approx 4.7\times10^{-14}$ erg s$^{-1}$ cm$^{-1}$ on day 234
(Deng et al. 2004), we find
the luminosity $L\approx 3.8\times10^{41}$ erg s$^{-1}$ which
significantly exceeds the upper limit found for $S/S_0=1$. This means that
either $S/S_0\gg1$ or the emission feature at 8500 \AA\ is
dominated by \Oi\ 8446 \AA\ as Deng at al. (2004) suggest. Below we
will see that the contribution of the \Oi\ 8446 \AA\ line is small so
the high luminosity of the \caii\ triplet is primarily
due to the large area ratio.

To show the effect of $Q_0$ we consider
the transparent case $Q_0=0$ and modestly opaque shell $Q_0=1$
for $\tau\gg1$.
The calculated profiles for a constant velocity shell with
$v_{\rm s}=6000$ km s$^{-1}$ and $\xi=0.05$ are shown
in Fig. \ref{f-single} (upper panel).
The finite optical depth smooths the boxy profile as expected.
In the limit of $Q_0\gg 1$, the profile is parabolic
with a weak asymmetry due to relativistic effects.
The relativistic effects arising from the Lorentz transformation of 
the frequency, wave vector and intensity are usually ignored in SN 
spectrum calculations. However as was demonstrated by Jeffery (1993) these
effects become noticeable even for expansion velocities 
of $\sim 10^4$ km s$^{-1}$ 
and generally result in the skewing of an emission component towards blue.

\subsubsection{Non-local scattering in SN ejecta}\label{sec-nonloc}

Although the contribution of the net emission from
the unshocked SN ejecta is small for SN2002ic
given the domination of the shock luminosity,
the SN ejecta can affect the line profile
via the scattering of line photon emitted by the CDS.
If the SN ejecta are optically thick in the same line, then
a line photon emitted by the shell inwards
can scatter on the unshocked \caii\  ions at the resonance point
with the distance from the emission point
\begin{equation}
s=R|\mu|\left(1-\frac{v_{\rm s}}{v_{\rm sn}} \right)\,,
\label{eq-nls}
\end{equation}
where $\mu=\cos\,\theta<0$. The resonance points reside on a
sphere of radius
$(1-v_{\rm s}/v_{\rm sn})R/2$ that contains the emitting point
($s=0$ for $\mu=0$).

To compute the effect of the non-local scattering of
line photons in SN ejecta we adopt the boundary velocity
of the SN ejecta $v_{\rm sn}=9000$ km s$^{-1}$; this value and
the shell velocity ($6000$ km s$^{-1}$) are
close to velocities in the standard model around day 230 (Fig. \ref{f-blc}).
Let $r_{\rm c}$ be a cut-off radius (in units of $R_{\rm s}$)
defined in such a way
that the Sobolev optical depth of the
SN ejecta $\tau_{\rm sn}=0$ in the outer zone,
$r>r_{\rm c}$, and $\tau_{\rm sn}>0$ in the inner zone, $r\leq r_{\rm c}$.
In both models the effect of non-local resonance scattering
is shown for two cases:  $r_{\rm c}=0.95$ and $r_{\rm c}=0.9$, assuming
$\tau_{\rm sn}=5$ in the unshocked SN (Fig. \ref{f-single}, lower panel).
Profiles demonstrate significant skewing towards the blue and
the emergence of a symmetric broad base produced by
scattered photons in the fast SN ejecta with velocity
$v_{\rm sn}>v_{\rm s}$.
The shape of the line profile depends on $r_{\rm c}$:
the skewing is larger for larger $r_{\rm c}$.
As we see below, the non-local scattering can account for
the asymmetry of the \caii\ doublet observed in SN~2002ic.

\subsection{Models of the \caii\ features}\label{sec-caprof}

We start with the modeling of the \caii\ doublet.
A Monte Carlo technique is used for the radiation transfer computations.
The parametrized emission rate of the CDS fragments in lines
of the \caii\ doublet is set assuming equipartition of the
upper levels. The non-local interaction in the shell is treated
as a pure absorption process, whereas the resonance
interaction of line photons with the unshocked SN ejecta is
considered as pure scattering.
The quasi-continuum radiation, presumably emitted by the shell can be
scattered by \caii\ lines of SN ejecta, whereas the scattering
in other lines (e.g., \feii) is neglected.
The quasi-continuum intensity, the slope of the emergent spectrum,
and the relative line intensity are free parameters.


\begin{table}
  \caption{Model parameters}
  \bigskip
  \begin{tabular}{lccccccc}
  \hline

Model & $Q_0$ &  $v_{\rm s}$ & $v_{\rm sn}$ & $r_{\rm c}$ & $\tau_{\rm sn}$
   & $C\mbox{(\Oi)}$  \\
    &   &  km s$^{-1}$  &  km s$^{-1}$  &    &           &     \\

\hline

DC0   &  0   &    5936      &  9092  & 0.95    &  3     &           \\
DC1   &  1   &    5936      &  9092  & 0.95    &  3     &           \\
DL0   &  0   &    5936      &  9092  & 0.95    &  3     &           \\
DL1   &  1   &    5936      &  9092  & 0.95    &  3     &            \\
TL0   &  0   &    6053      &  9230  & 0.95    &  0.005  &   0.25  \\
TL0   &  1   &    6053      &  9230  & 0.95    &  0.005  &   0.25    \\

\hline
\end{tabular}
\label{t-par}
\end{table}

\begin{figure}
  \centering
  \vspace{9 cm}
  \includegraphics{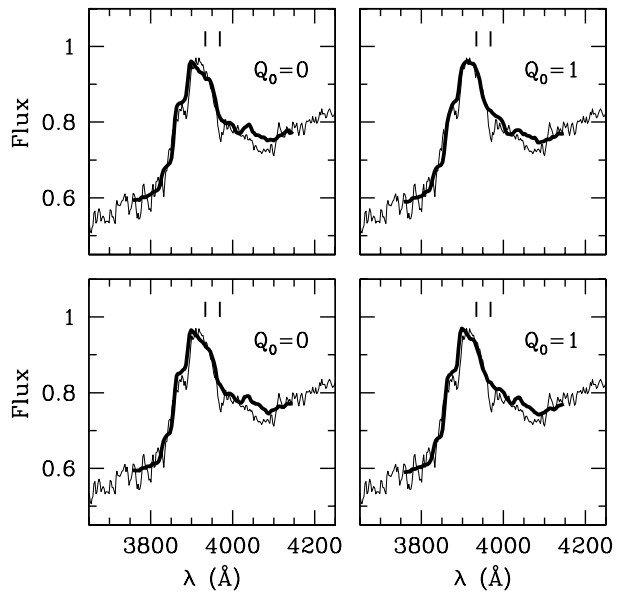}
  \caption[]{Calculated profiles of the \caii\ 3934, 3968 \AA\
  doublet ({\em thick} line) overplotted on the spectrum on day 244
  (Wang et al. 2004). Upper panels show models with constant velocity
  in the shell and different $Q_0$ (model DC0, left, and model
  DC1, right). The lower panels are the same except for
  the linear velocity distribution in the shell (model DL0, left,
   and model DL1, right).
  Two vertical bars show the rest frame positions of 3934 \AA\ and
  3968 \AA\ lines.
       }
  \label{f-cad}
  \end{figure}

The calculated  \caii\ doublet assuming a constant velocity and
an occultation optical depth
$Q_0=0$ (model DC0) and $Q_0=1$ (model DC1), and assuming
a linear velocity distribution with the same $Q_0$ values
(models DL0 and DL1, respectively) are
shown in Fig. \ref{f-cad} overplotted on the spectrum
on day 244 (Wang et al. 2004).
The computed spectra are smoothed by a gaussian filter with
FWHM=7 \AA\ to take into account the finite spectral resolution.
Velocities $v_{\rm s}$ and $v_{\rm sn}$ of
the standard interaction model as well as other parameters are
presented in Table 1. The table gives also
$Q_0$, the optimal cut-off radius ($r_{\rm c}$) and the optical depth
of the SN ejecta in the \caii\ 3934 \AA\ line  ($\tau_{\rm sn}$).
We found that efficient scattering in the SN
ejecta is needed with $\tau_{\rm sn}\geq3$ in \caii\ 3934 \AA.
The optimal cut-off radius for all the models,
$r_{\rm c}=0.95$, coincides with the inner boundary of the
line-emitting shell.

All four models shown in Fig. \ref{f-cad} reproduce the major
properties of the observed \caii\
doublet, particularly, the width, shape and the
skewing towards the blue. The deviation in the red wing may be related to
the behavior of the underlying quasi-continuum in the observed spectrum.
Models with constant and linear velocity laws in the shell
are practically indistinguishable. We will use below only the linear
law model that presumably mimics the possible random component of the
velocity field in the shell.

\begin{figure}
  \centering
  \vspace{5.5 cm}
  \includegraphics{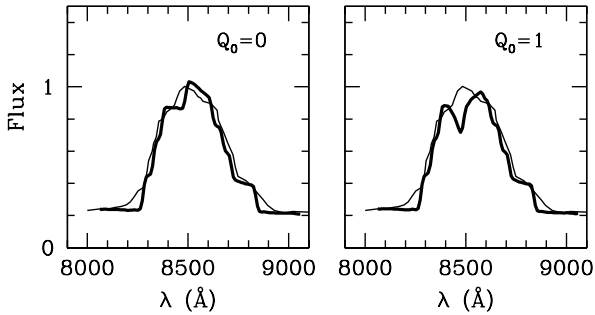}
  \caption[]{Calculated profile of the \caii\ infrared triplet
  blended with \Oi\ 8446 \AA\  for the models
  TL0 (left) and TL1 (right)
   overplotted on the observed profile on day 234 (Deng et al. 2004).
         }
  \label{f-cat}
  \end{figure}

Using models DL0 and DL1, we also simulated the \caii\ triplet
(Fig. \ref{f-cat}). The model profiles are overplotted on the
observed profile in the Subaru spectrum on day 234 (Deng et al. 2004).
Parameters of the corresponding models (TL0 and TL1) including velocities of
the standard model are
given in Table 1 together with the optimal contribution of the
\Oi\ 8446 \AA\ line to the 8500 \AA\ feature $C\mbox{(\Oi)}$.
The overall fit of the model TL0 is satisfactory.
The derived contribution of the \Oi\ 8446 \AA\ line to the 8500 \AA\ feature
$C\mbox{(\Oi)}=0.25$ is constrained with an uncertainty of
$\pm 0.05$.
However, model TL1 with $Q_0=1$ shows a pronounced central absorption
feature caused by the resonance non-local absorption in the
shell. This feature is inconsistent with
the observed profile. We conclude that the fragmented shell
should be characterized by a low probability of non-local
radiation coupling in the \caii\ triplet. We will return to this
issue below (Sec. 3.3).

The observed 8500 \AA\ profile is slightly broader than the model feature,
especially in the wings. However, one must be
cautious about attributing the wings to the \caii\ triplet because
the quasi-continuum also contributes to the 8500 \AA\ feature,
as will be demonstrated below.
Given the uncertainties of the behavior of the quasi-continuum
and the satisfactory fit of the \caii\ doublet and triplet,
we conclude that the \caii\ features are consistent with
a spherical model and expansion velocities predicted by
the interaction model.

We performed computations with velocities multiplied by a factor larger
than unity. We found that slightly larger expansion
velocities (by $\sim 5$\%) produce  better fits to the 8500 \AA\ features.  
However,
this conclusion is uncertain because of a possible
contribution from the quasi-continuum.

\subsection{CDS temperature and $S/S_0$ ratio}\label{sec-area}

The contribution of the \Oi\ 8446 \AA\ line to the 8500 \AA\ feature found 
on day 234 ($\sim 25$\%), combined with the total flux
$\approx 4.7\times10^{-14}$ erg s$^{-1}$ cm$^{-2}$ (Deng et al. 2004),
implies a flux in the  \Oi\ 8446 \AA\ line 
$\approx 1.2\times 10^{-14}$ erg s$^{-1}$ cm$^{-2}$, and
$\approx 3.5\times10^{-14}$ erg s$^{-1}$ cm$^{-2}$ in
\caii\ triplet.
Given the \caii\ doublet flux $\approx 10^{-14}$ erg s$^{-1}$ cm$^{-2}$
in the same spectrum (Deng et al 2004), one gets the
triplet-to-doublet ratio $F(8579)/F(3945) \approx 3.5$. Assuming that
both \caii\ excited terms ($^2$D and $^2$P) have
the same excitation temperature
and using the scaling $F\propto \nu B_{\nu}(T)$ from 
the ratio $F(8579)/F(3945)\approx3.5$ we obtain $T\approx 4430$ K.

The absolute luminosities of the \caii\ features can be used to
estimate $T$ and the area ratio $S/S_0$ simultaneously
(see equation [\ref{eq-lumir}]). To make the procedure
reliable, we abandoned the assumption that the excitation temperature for
both excited terms is the same.
Instead, the statistical equilibrium for a three level \caii\ 
model is solved with collisional and radiative
transitions taken into account. The radiation transfer in a fragment
is treated in the static average escape probability approximation.
The adopted Ca abundance is 0.03 by mass, the same as used in
Section \ref{sec-shell}.
The ionization fraction of \caii\ is calculated in the LTE
approximation, which implies that \caii\ is the dominant stage
for a wide range of
temperature,  $4000-6000$~K.
The solution of statistical equilibrium then yields the excitation
temperature in both \caii\ transitions as a function of the
electron temperature $T_{\rm e}$ and $S/S_0$, which permits us
to calculate the luminosities of the \caii\ features
(equation [\ref{eq-lumir}]) and to determine
$T_{\rm e}$ and $S/S_0$ from a comparison with the
observed luminosities.

\begin{figure}
  \centering
  \vspace{6.0 cm}
  \includegraphics{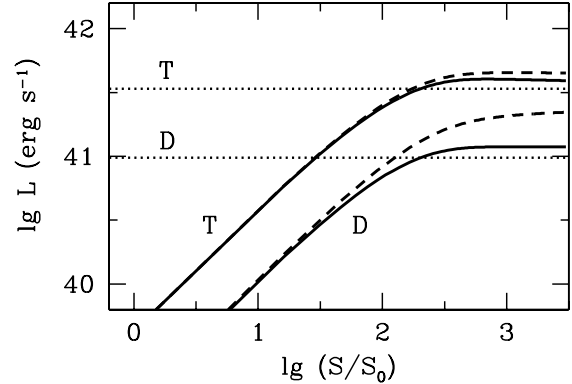}
  \caption[]{Calculated luminosities of the \caii\ doublet and
   triplet on day 234 as a function of the area ratio for
    $T_{\rm e}=4430$ K with zero albedo ({\em thick} line) and
   finite albedo ({\em dashed} line) of dense fragments.
     The observed luminosities of the \caii\ doublet and infrared triplet
   are plotted by {\em  dotted} lines. Subscripts T and D
   refer to triplet and doublet feature.
       }
  \label{f-lums}
  \end{figure}

The \caii\ levels  are found to be well thermalized, so 
to reproduce the observed luminosities of \caii\ features we need 
only two parameters: $T_{\rm e}$ and $S/S_0$, assuming $\xi=0.05$.
Figure \ref{f-lums} shows
the behavior of the luminosities of the \caii\ features as a function
of $S/S_0$ for the optimal temperature $T_{\rm e}=4430$ K.
The plot clearly demonstrates that the observed luminosities of the \caii\ features
 can only be reproduced for a large area ratio, $S/S_0\sim 10^2$.
This result provides strong evidence that the dense material
of the shocked ejecta experienced substantial deformation, fragmentation
and mixing in the intershock layer.

We studied the sensitivity of the luminosity of \caii\ lines 
to the adopted mass locked in line-emitting fragments 
($M_{\rm e}$) and found 
that in the range of $S/S_0<300$ the luminosity of both \caii\ 
multiplets are well reproduced for $M_{\rm e}\geq 10^{-2}~M_{\odot}$
and cannot be reproduced for smaller mass, $M_{\rm e}<10^{-2}~M_{\odot}$.
This indicates that the amount of 
 the Ca-rich matter locked in the \caii\ line-emitting gas is 
 $\geq0.01~M_{\odot}$ (for the mass fraction of Ca $\approx 0.03$). 
 
The conclusion that the area ratio is as large as $\sim 10^2$ 
raises a serious problem. The
local occultation optical depth of the shell with an area ratio
$S/S_0\sim 10^2$ is $Q_0\sim 1$. This seems
inconsistent with the  claim (Sec. 3.2)
that the non-local radiative coupling in the
\caii\ triplet is weak. For the spherical shell, this requirement
implies $Q_0\ll 1$.
In an attempt to resolve the inconsistency, we considered
a model of the \caii\ feature formation with a finite
albedo $\omega$ of dense fragments for the incident line photons.
To determine $\omega$ we performed a Monte-Carlo simulation of
the history of the
incident \caii\ doublet and triplet photons assuming a three
level model of \caii.
The line luminosities for the finite albedo are shown in
Figure \ref{f-lums}. The required value of the area ratio
in the new model
is only slightly lower compared with the previous model,
so the problem of large $Q_0$ remains unsettled.

A way out might be in the assumption that
the accepted model of the fragmented shell is too simple to
predict the upper
limit for the occultation optical depth from the \caii\ triplet. For
instance, the large scale
corrugation of the shell and/or patchy structure of the line-emitting
zones might result in voids lacking
the absorbing dense gas at the resonance points. As a result,
the large local value of $Q_0$ may be reconciled with
the low absorption probability for the non-local interaction
in the \caii\ triplet.

\subsection{Quasi-continuum model} \label{sec-cont}

We now check our conjecture that
the quasi-continuum of SN~2002ic is produced by 
the emission of the fragmented cool shocked ejecta in numerous
lines of Fe-peak elements. To this end, we compute the
emergent spectrum assuming a model of the fragmented  shell
width $\Delta R/R=0.1$ for two values of the area ratio $S/S_0=50$
and $S/S_0=1$. The shell patchiness indicated by the \caii\
triplet modelling will be described by the surface (and volume)
filling factor $f_{\rm qc}$ of quasi-continuum emitters.
It should be emphasized that the filling factor refers only to the
volume fraction of mixture and not to the volume
fraction of the dense material that is, of course, much lower.
The radiation transfer in the shell will be computed
assuming a spherical shell, with two corrections.
First, the
emitting volume will be multiplied by the factor $f_{\rm qc}$, and, second,
we take into account a finite probability of the
photon escape through the holes of the opposite side of shell,
$w=|\mu|(1-f_{\rm qc})$. Here $\mu$, the cosine between
the photon wave vector and the radius, describes approximately
the angular dependence of $w$. The unshocked ejecta will
be considered as transparent and the net emission of this ejecta
will be neglected. We also omit the presence of a shocked H-rich cool 
gas in the intershock layer.

We consider the seven most abundant Fe-peak elements,
Ti, V, Cr, Mn, Fe, Co, Ni, assuming their relative abundance to be solar.
The latter is a sensible approximation for the explosive nucleosynthesis 
in the full incineration regime. 
Doublet and triplet lines of \caii\ are also included in the
line list.
The initial line list is extracted from the VALD data base (Kupka et al. 1999)
assuming conditions in a red giant atmosphere with
$T_{\rm eff}=5000$ K and gravitational acceleration $\log\,g=1.5$.
The list contains about 19000 lines of neutral and singly ionized species
with a line optical depth $>10^{-5}$ in the range of $3500-10000$ \AA.
Using this list, we compile the input line list for a particular density,
assuming Saha ionization with a temperature $T_{\rm ion}$ and
Boltzmann excitation with a temperature $T_{\rm ex}$.
The line optical depth is restricted by the condition $\tau>10^{-4}$.
For the relevant conditions, the most abundant species are singly ionized
($\phi_2\approx 1$), while the ionization fraction of neutrals
is $\phi_1\sim 10^{-2}$ and that of double ions is $\phi_3\sim 10^{-6}$.

The radiation transfer is calculated using a Monte Carlo technique.
The photon interaction with
a dense fragment is treated as a partial absorption with an
albedo $\omega$. We adopt $\omega=0.1$ for all the lines,
although we found that the result is practically the same in the case of
pure absorption ($\omega=0$).
Scattering is assumed to be isotropic for the back and forward
scattering with a diffusion reflection probability $\tau/(1+\tau)$.
The computed spectrum is reddened using $E(B-V)=0.06$ (Deng et al. 2004).

\begin{figure}
  \centering
  \vspace{10 cm}
  \includegraphics{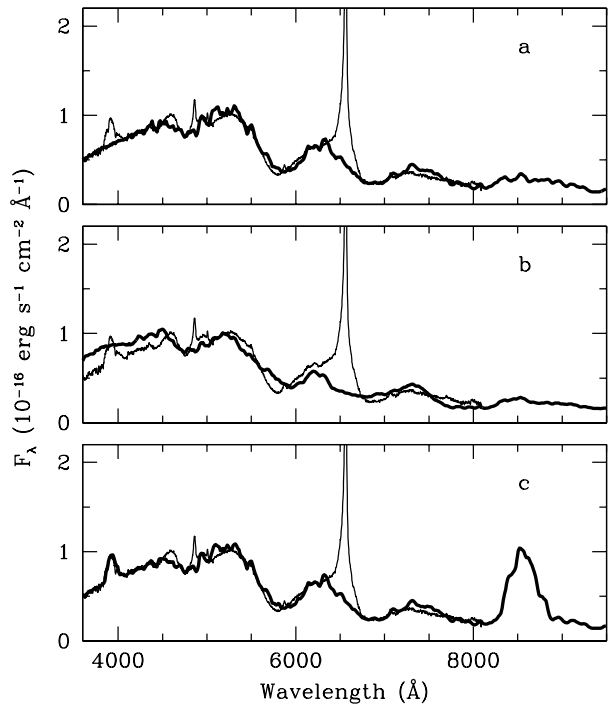}
  \caption[]{Synthetic spectra for the model of a fragmented
  shell overplotted on the observed spectrum of SN~2002ic on day 244
  (Wang et al. 2004). Panel {\em a} shows the model with a large area
  ratio ($S/S_0=50$, see text), whereas panel {\em b} shows a model with
  $S/S_0=1$. Panel {\em c} shows the same model as in panel
  {\em a} but with the addition of a \caii\ line-emitting zone.
         }
  \label{f-qcon}
  \end{figure}

An extensive search in the
parameter space ($T_{\rm ion}$, $T_{\rm ex}$, $f_{\rm qc}$)
for the model with $S/S_0=50$ 
led us to conclude that a best fit of the quasi-continuum in the spectrum
on day 244 is attained for the
parameters $T_{\rm ion}=6000$ K, $T_{\rm ex}=4300$ K, and
$f_{\rm qc}=0.125$ (Fig. \ref{f-qcon}a).
Note, the result is not very much sensitive to the area ratio; 
we adopt $S/S_0=50$, not $100$ that follows from \caii\ line analysis, 
because the latter value leads to a too large occultation optical depth 
(see discussion in Section \ref{sec-area}).
On the other hand, the optimal model with the small area ratio, 
$S/S_0=1$ and $f_{\rm qc}=1$ (Fig. \ref{f-qcon}b) seems to be inconsistent 
with the observed shape of the quasi-continuum. This implies that
the model with a large area ratio is preferred thus supporting 
the conclusion made from \caii\ line analysis.

The fit is good despite the fact that the quasi-continuum model 
is rudimentary;
the number of free parameters is small.
The agreement suggests that we are on the right track in the
interpretation of the quasi-continuum.
This result makes the identification of
\oi, \oii, \cafii\ lines for the 6400 \AA\ and 7300 \AA\ bumps
(cf. Deng et al. 2004) unneccessary. Otherwise, there would be serious
problems in the interpretation of the identified emission lines.
Another remarkable fact is that the quasi-continuum
contributes to the 8500 \AA\
emission feature thus alleviating the problem of the origin of
broad wings in the \caii/ \Oi\ 8500 \AA\ feature.
We checked the contribution of different ions to the 
quasi-continuum and found that, as expected, the \feii\ lines 
are dominant. This elucidates the role of the ionization 
temperature as a fitting parameter that regulates
 the contribution of neutral species, \fei\ in particular.
The quasi-continuum in the region 
$\lambda < 5500$ \AA\ is very reminiscent to that in early 
SN~1988Z, which also has been identified with \feii\ lines 
(Chugai \& Danziger 1994). 

Interestingly, the smooth quasi-continuum in the range $\lambda< 4500$ \AA\
(Fig. \ref{f-qcon}a) is the result of 
radiation saturation at the Planck intensity due
to the high density of lines in this band that results in
the large optical depth of the shell, viz.
$\tau_{\rm shell}\sim\langle Q\rangle N_{\rm line} >1$,
where $N_{\rm line}$ is the number of lines in the characteristic
wavelength
interval $\lambda(v_{\rm s}/c)(\Delta R/R)$ with the average
local occultation optical depth $\langle Q\rangle$.
The saturation effect is absent
in a model with small area ratio $S/S_0\sim 1$
(Fig. \ref{f-qcon}b). Because  the line density in the ultraviolet (UV) 
region is very high, we expect that the UV quasi-continuum should retain a 
smooth black-body appearance with $T\sim4300$~K in the short wavelength
 region ($\lambda<3500$~\AA).

As shown in Fig. \ref{f-qcon}a,
the quasi-continuum model is unable to produce strong \caii\ emission features.
This is related to the fact that the quasi-continuum is a
superposition of a large number of weak lines affected additionally by 
the line overlapping. We propose, therefore, that
the \caii\ emission and quasi-continuum should form in
{\em different} zones of the shocked ejecta. These zones, 
presumably, correspond to  Fe-poor (Ca-rich) gas and 
 Fe-rich gas, respectively. 
The obvious physical reason for these distinctive zones  is 
the abundance stratification of SNe~Ia with the inner 
layers incinerated up to Fe-peak elements and the
outer layers having undergone incomplete burning. 
In this situation one expects that 
the outer Fe-poor gas that has passed through the reverse shock first 
has enough time to spread over the large volume of the intershock 
layer, whereas the inner Fe-rich material shocked recently occupies only a
small fraction of the intershock layer.

To illustrate the proposed scenario for the \caii\ lines,
we consider a combined model of
the fragmented shell: apart from the quasi-continuum line-emitting
zones suggested by the model with $f_{\rm qc}=0.125$,
we consider the \caii\ line-emitting zones
with a filling factor $C(1-f_{\rm qc})$, where $C$
is a fitting parameter. We assume a Ca abundance
$\approx 0.03$ by mass (the result only weakly sensitive to Ca abundance)
and neglect the Fe-peak elements in the \caii\ line-emitting zone.
For \caii\ lines, the scattering in the unshocked SN ejecta on \caii\
is taken into account as described in the previous section.
The same area ratio $S/S_0=50$ and $T_{\rm ex}=4300$ K are adopted
in the \caii\ and quasi-continuum zones.
We omit here the contribution of \Oi\ 8446 \AA. 

The result presented in Fig. \ref{f-qcon}c is obtained for
$C=0.8$, which suggests that $\sim 70$\% of the fragmented
shell volume is occupied by Fe-poor \caii\ line-emitting
material. The filling factor again refers only to the
volume fraction of the mixture and not the dense material.
Although we do not intend to describe
the \caii\ profiles and intensities in detail,
the combined model provides a satisfactory representation of the
overall spectrum, except for \ha\ and \hb. The modelling thus 
confirms the proposed scenario based upon the expected 
abundance stratification in the outer layers of the SN ejecta.

We now address the question of how low the abundance of Fe-peak 
elements in the \caii\ line-emitting matter can be and retain sufficient 
contrast between the quasi-continuum and
the \caii\ lines, especially the \caii\ doublet that resides in 
the line-rich spectral range. The answer depends on the 
assumed mass locked in the  \caii\ line-emitting dense gas ($M_{\rm e}$). 
We adopt $M_{\rm e}\approx 0.02~M_{\odot}$ and a Ca abundance 0.03, which 
is close to the admissible lower limit ($M_{\rm e}\geq 0.02~M_{\odot}$)
and is consistent with the mass and composition of the Fe-poor Ca-rich
layer produced by the delayed detonation model (Iwamoto et al. 1999). 
For these parameters and the values of $T_{\rm ion}$
and $T_{\rm ex}$ adopted above 
for the case of $S/S_0=50$,  we found that the mass fraction of 
Fe-peak elements should 
not exceed $2\times10^{-4}$ or $\sim0.1$ of solar abundance.
This upper limit indicates a low 
primordial metallicity of the pre-SN and, possibly, of the host galaxy 
as well. 

To summarize, the major observational properties of the SN~2002ic spectrum
at an age of $\sim 240$ days may be explained in the 
framework of a model of a spherical, locally inhomogeneous shell with
an expansion velocity predicted by the standard hydrodynamic model.

\section{Discussion and conclusions}\label{sec-conc}

The aim of this paper was to develop a unified model 
for the light curve and spectrum of SN~2002ic assuming approximate spherical
symmetry.
Our principal result is that the \caii\ line profiles and the quasi-continuum
are consistent with the interaction model provided that
the SN~Ia has the maximum possible kinetic energy.

The high explosion energy is in line  with the
high peak absolute magnitude of SN~2002ic.
According to Hamuy et al. (2003), at light maximum SN~2002ic
was as bright as $M(V)\approx -20.3$ ($H_0=65$ km s$^{-1}$ Mpc$^{-1}$).
Given the contribution of the veiling
continuum of about $0.3$ mag on Dec. 3 in $V$ band ($5000-6000$ \AA)
(Hamuy et al. 2003), we conclude that the peak absolute magnitude of the SN~Ia
was  $M(V)\approx -20$, close to that of extremely bright SN~Ia, SN~1991T 
($M(V)\approx -20.2$, cf. Fisher et al. 1999).   
The magnitude at the maximum coincides also with that predicted 
by the detonation model DET1 (H\"{o}flich \& Khokhlov 1996).
This coincidence is another justification for our choice of the
high kinetic energy for the SN~2002ic event. Yet, as we noted above, the 
strong \caii\ emission lines indicate that 
the explosion should be different from the DET1 model in the sense that 
the degree of incineration in the outer layers should be lower to produce 
enough ($\geq10^{-2}~M_{\odot}$) Ca-rich Fe-poor material. 

Modeling of the spectra of the
fragmented cool dense shell formed in the intershock layer provides
strong evidence that the quasi-continuum and \caii\ lines
are emitted by different zones. This presumably
reflects the stratification of the composition of the SN ejecta
in the sense that the outer layers responsible for
the \caii\ lines are Fe-poor, while the inner layers
responsible for the quasi-continuum are Fe-rich. The Fe-poor ejecta
are shocked earlier and, therefore, spread throughout the
intershock layer, while Fe-rich ejecta are shocked later and thus
have not yet mixed with the Fe-poor material at $t\sim 240$ d.
This picture predicts that at a later
epoch ($t>240$ d) both Fe-poor and Fe-rich
shocked components  should be mixed with each other.
As a result, the \caii\ doublet must disappear at a later
epoch due to the mixing of the \caii\ line-emitting gas
with optically thick quasi-continuum emitters.
SN~1999E
showed a gradual disappearance of the \caii\ 3945 \AA\ doublet
in  spectra between days 139 and 361 (Rigon et al. 2003),
 in accord with these expectations.

The high luminosity of
the \caii\ doublet and triplet in SN~2002ic on day 234 indicates that
the dense shocked ejecta form a structure with a
large area ratio, $S/S_0\sim10^2$. A high area ratio ($S/S_0\sim 50$)
is also preferred by the quasi-continuum model.
The area ratio, which is common in the treatment of the physics
of mixing fluids (e.g., Catrakis et al. 2002), has not been used before
in the interpretation of optically thick emission lines from an
intershock layer.
It was recognized earlier that the interpretation of the 
X-ray and radio emission of an interacting SN depends on the
presence of a mixing layer, while the \ha\ emitted by the
mixed dense material may reveal the fine structure related to the RT spikes
(Chevalier \& Blondin 1995).
Here we add to the list of effects of the mixing layer
the crucial role of the large area ratio for
the interpretation of the \caii\ lines and quasi-continuum
in SN~2002ic.

The fact that the area ratio in SN~2002ic on day 234 is
large, $S/S_0\sim 10^2$, suggests that we observe an advanced
stage of the mixing of dense shocked ejecta in the intershock zone.
Experiments on the mixing of fluids show
that the interface between fluids reveals fractal behavior.
Specifically, the cumulative area of the interface
above the scale $\lambda$ is $S\propto \lambda^{-\alpha}$,
where $\alpha\approx 0.35$
(Sreenivasan et al. 1989). If this law also holds  in the case of the
mixing of the dense shocked ejecta in the intershock layer of SN~2002ic,
then the dense gas should produce structures with
scales as small as $\lambda/R \propto (S/S_0)^{-1/\alpha} \sim 10^{-6}$.
The minimum linear scale that corresponds to the thickness
of the dense fragments should be  even smaller,
$\sim\delta(S_0/S)\sim 10^{-8}R$,
where $\delta$ is the original thickness of the CDS
(Section \ref{sec-mech}).
It would be interesting to check the possibility
of creating the mixing layer in SN~2002ic with
$S/S_0\sim 10^2$ in numerical experiments using high
resolution 3D hydrodynamical simulations, although it seems to be
beyond the capabilities of present-day computers.
To estimate the area ratio from a numerical
experiment, one needs to find the average number of intersections of dense
peaks with the radius in the intershock layer $N_{\rm is}$. The
area ratio is then $S/S_0\approx 2N_{\rm is}$.

We do not yet fully understand all the mechanisms responsible for the
major spectral features seen in the late emission from SN~2002ic.
In addition to the \ha\ line, an important issue ignored in our study is the
problem of the \Oi\ 8446 \AA\ emission, 
although it is not a dominant constituent of the 8500 \AA\ bump anymore.
The oxygen excitation mechanism should predict the
absence of strong \Oi\ 7774 \AA\ in SN~2002ic.
The fluorescence pumping of \Oi\ 8446 \AA\ in the \ha-emitting gas
(Deng et al. 2004) meets this
requirement. This mechanism predicts a 
comparable intensity of \Oi\ 11287 \AA. The absence of this line 
in the spectrum of SN~1999E around $\sim200$ d 
(Rigon et al. 2003, their Fig. 6) casts doubt on the
fluorescence mechanism. An alternative possibility is that
the \Oi\ 8446 \AA\ line is emitted by the oxygen-rich matter
in the outermost layers of the SN ejecta due to non-thermal
collisional excitation by secondary
electrons that are created by high energy photoelectrons.
Excitation by secondary electrons also yields a high
 \Oi\ 8446 \AA/\Oi\ 7774 \AA\ intensity ratio (Stolarski \& Green 1967).
An interesting possibility that Ly$\alpha$ pumping line 
of \feii\ 8451 \AA\ might contribute to 8500 \AA\ band instead of 
\Oi\ 8446 \AA, as was the case, possibly, for SN~1995N 
(Fransson et al. 2000), seems unlikely here because of the absence 
of another strong Ly$\alpha$ fluorescent band of \feii\ at 
$\sim 9150$ \AA.
Despite of a possible presence of a large amount of oxygen in the 
cool shocked SN gas (up to  $\sim 0.1~M_{\odot}$), the strong 
\oi\ line is not expected. Adopting Boltzmann population for 
$T=4400$~K and oxygen mass of $0.1~M_{\odot}$ one gets the 
luminosity of this line of about $\sim10^{39}$ erg s$^{-1}$, i.e., 
a factor of $\sim10^{-2}$ lower compared to the luminosity of the 
observed "6300 \AA" emission band on day 244.

In view of the possibility that our hydrodynamic model may
underestimate the expansion velocity of the SN/CSM
interface by about $5$\%, we discuss the implications
for our model.
The emergent X-ray luminosity
from the forward shock in the standard model is comparable or even exceeds
the optical bolometric luminosity after  maximum light 
(see Fig. \ref{f-blc} and Fig. \ref{f-xlc}). Hence, there
is an extra energy
reservoir which might, in principal, be transformed into the optical
band under the proper conditions.
This possibility could be realized in a model of  interaction
with a clumpy CSM. As a result of (i) softer
X-rays from cloud shocks than for a smooth CSM, and 
(ii) the larger geometrical probability of absorption of
X-rays from clouds penetrating the SN ejecta, the optical
output from the interaction should be larger.
Therefore, the same bolometric luminosity can be produced for a lower
average CS density and
higher expansion velocity of the SN/CSM interface compared with a smooth CSM.

The spherical geometry of the CS interaction of SN~2002ic,
advocated here, finds additional support
from statistical arguments already mentioned in 
the Introduction. We know  three
events of SNe~Ia in a dense CSM (SN~1997cy, SN~1999E, SN~2002ic)
and they show similar widths and shapes of spectral features at
a similar phase (see Deng et al. 2004, their Fig. 1). This fact
suggests that a bi-polar structure of the SN/CSM interface with high aspect
ratio is unlikely.

At first glance the picture of  spherical interaction is
incompatible with the detection of  intrinsic
polarization of SN~2002ic radiation at a level of $\sim 0.8$ \%
(Wang et al. 2004). Simple
estimates show that the Thomson scattering in a moderately aspherical
CS envelope
with $\tau_{\rm T}\approx 0.1$ is not able to produce such a high
polarization.
The issue may be resolved in a model in which
the polarization is acquired in the
scattering of the SN radiation on the dust in a distant
($\geq10^{17}$ cm) aspherical CSM.

The resemblance of the light curves for all three SN~2002ic-like
events (see Deng et al. 2004) suggests a similar regime of
 mass-loss from their presupernovae
and a common origin of the progenitors. As mentioned before
(Chugai \& Yungelson 2004), the uniformity lends support for the SN\,1.5
evolutionary scenario suggested by Iben \& Renzini (1983). 
Yet we do not rule out the binary scenario. 
In this case, the WD companion should be a supergiant 
with the mass $>2~M_{\odot}$ (the CS envelope, 
$\approx1.6~M_{\odot}$, plus the supergiant core, $>0.5~M_\odot$).
In the binary scenario, there is no clear reason for the supergiant 
to have experienced strong mass loss just before the SN explosion. 
The synchronization of the vigorous mass loss 
and the SN event might be a more natural outcome in the single 
AGB star scenario with the C/O core approaching the Chandrasekhar limit.

A general belief is that Chandrasekhar mass white dwarfs
cannot be produced from intermediate-mass single stars because 
the growth of the C/O core during the thermal pulsation stage
is prevented by the removal of the hydrogen envelope 
(Bl\"{o}cker 1995, and references therein). In this regard, 
it is highly remarkable that at least two supernovae 
(SN~2002ic and SN~1997cy) exploded in dwarf galaxies, 
as already noted by Chugai \& Yungelson (2004).
 Dwarf galaxies are characterized by low metallicity, which thus might
explain why the CO core of an asymptotic giant branch star is successful in
attaining the Chandrasekhar limit before the hydrogen envelope
is lost (Chugai \& Yungelson 2004). 
An evidence for the low metallicity of the presupernova of SN~2002ic 
comes from our modelling of the \caii\ lines and quasi-continuum 
(Sec. \ref{sec-cont}). 
The possibility of SN\,1.5 explosions in 
metal-poor galaxies was suggested also by Zijlstra (2004).
It would be  interesting to check whether all the host galaxies
of SN~2002ic-like supernovae are actually metal deficient.
If the suggested link between SN~2002ic-like 
events and low metallicity is correct, then their occurrence rate relative to 
normal SN~Ia must increase with the redshift for $z>2-3$, when 
the average cosmic metallicity was lower than solar.

If SN~2002ic-like events stem from SN~1.5 explosions, their
low fraction of SN~Ia, $<1$\%, can be used to estimate the
the fraction of white dwarfs created with a mass similar to
the Chandrasekhar mass. If only $\sim 1$\% of all WDs can become
SN~Ia, this suggests that only $\lsim10^{-4}$ of white dwarfs 
in our Galaxy were formed with a mass close to the Chandrasekhar 
mass. Since these white dwarfs are likely to have formed in a 
metal poor environment, they are probably very old 
(age $\sim 10^{10}$ yr) and thus of low luminosity,
$<3\times10^{-5}~L_{\odot}$ (Prada Moroni \& Straniero 2003). 
The number of known cool white dwarfs is low, $<40$ (Salim et al. 2004),
and will probably remain too low for the nearest future
to test the idea whether high mass white dwarfs were produced 
during early epochs when the metallicity was low.

The possible link between SN~2002ic-like supernovae and
hypothetical SN\,1.5 events, combined with arguments in favour
of a high energy SN~Ia event in this case,
raises the  question of whether conditions in a
hot Chandrasekhar mass C/O WD that just got rid of the
remnants of the hydrogen envelope favour a WD explosion
with a high kinetic energy of ejecta. 

The SN\,1.5 origin of SN~2002ic suggests that the initial 
mass at the main sequence is $\sim8~M_{\odot}$. 
With the Chandrasekhar core and $1.6~M_{\odot}$ in the 
dense CS envelope we expect that about $5~M_{\odot}$ had to 
be lost at the previous AGB stage. This means that outside 
the dense CS envelope ($r>3\times10^{16}$ cm) 
there should be a dense red supergiant 
wind with a velocity of the order $\sim 10$ km s$^{-1}$ and a
mass loss rate of $\sim 10^{-4}~M_{\odot}$ yr$^{-1}$. 
Keeping in mind the analogy with SN~1979C, this wind may be revealed 
in the infrared (IR) as an IR echo (Dwek 1983), and in the radio 
as a result of CS interaction (Chevalier 1982a, Weiler et al. 1986).
However, the radio emission from the distant SN~2002ic ($\sim 285$ Mpc) 
may pose a difficult challenge to detect. If the radio luminosity is 
comparable to that of SN~1979C (Weiler et al. 1986) then 
the expected radio flux from SN~2002ic around 1000 day should be of 
the order of $\sim 0.03$ mJy in the $6-20$ cm band, which is close
to the detection limit of VLA. A more promising target for the 
detection of the late time interaction of SN~2002ic-like 
supernovae with the presumed AGB wind is SN~1999E at a
distance of $\sim 112$ Mpc, or any closer SN~2002ic-like event that
will be discovered in the future. Note, that at the epoch of 
the interaction with the dense CS shell ($t<500$ d), the radio 
emission is severely attenuated by free-free absorption
at wavelengths $\lambda> 1$ cm. With the parameters used in
our models, and adopting temperatures for the shell and wind
in accordance with the models of Lundqvist \& Fransson (1988), we
find that free-free absorption is a likely reason why the 
observations of Berger \& Soderberg (2003) and Stockdale et al. 
(2003) did not detect radio emission from SN~2002ic.

A better understanding of the phenomenon of SN~2002ic-like
events requires an accurate energy audit, for which
the direct detection of the  X-ray emission 
as well as IR measurements are important. 
SN~2002ic-like events at an age of $200-500$ days 
are excellent targets for Chandra, XMM and INTEGRAL.

\section{Acknowledgements}
We are grateful to Lifan Wang for sharing his spectrum
of SN~2002ic and Tanya Ryabchikova for the help in
retrieving the line list from VALD.
NNC received partial support from RFBR grant 02-17255
and RAC from NSF grant AST-0307366. Support from the Royal Swedish 
Academy is also acknowledged. PL is further supported by the 
Swedish Research Council.

{}


\begin{thebibliography}{}

\bibitem{} Arnett W.D., 1980, ApJ, 237, 541
\bibitem{} Bl\"{o}cker T., 1995, A\&A, 297, 727
\bibitem{} Berger T., Soderberg A.M., 2003, IAUC No. 8157
\bibitem{} Blondin J.M. 2001, in  S.S. Holt and U. Hwang eds.
Young Supernova Remnants. Eleventh Astrophysics Conf., AIP, p.59
\bibitem{} Blondin J.M,  Marks B.S., 1996, NewA, 1, 235
\bibitem{} Blondin J.M., Ellison D.C. 2001, ApJ, 560, 244
\bibitem{} Catrakis H.J., Agguirre R.C., Ruiz-Plancarte J. 2002,
J. Fluid. Mech., 462, 245
\bibitem{} Chevalier R.A., 1982a, ApJ, 258, 790
\bibitem{} Chevalier R.A., 1982b, ApJ, 259, 302
\bibitem{} Chevalier R.A., Blondin J.M., 1995, ApJ, 444, 312
\bibitem{} Chevalier, R.A., Fransson, C., 1994, ApJ, 420, 268
\bibitem{} Chugai N.N., 1992, Sov. Astron., 36, 63 
\bibitem{} Chugai N.N., Danziger I.J., 1994, MNRAS, 268, 173
\bibitem{} Chugai N.N. 2001, MNRAS, 326, 1448
\bibitem{} Chugai N.N., Yungelson L.R., 2004, Astron. Lett. 30, 65
\bibitem{} Cid Fernandes R.C., Terlevich R., 1994, in Tenorio-Tagle G.
ed. Violent Star Formation. From S Doradus to QSO. CUP, Cambridge, p. 365
\bibitem{} Deng J., Kawabata K.S., Ohyama Y., Nomoto K., Mazzali P.A.,
Wang L., Jeffery D.J., Iye M., 2004, ApJ, 605, L37
\bibitem{} Dwarkadas V.V., Chevalier R.A., 1998, ApJ, 497, 807
\bibitem{} Dwek E., 1983, ApJ, 274, 175
\bibitem{} Fisher A., Branch D., Hatano K., Baron E., 1999, MNRAS, 304, 67
\bibitem{} Fransson C. et al., 2002, ApJ, 572, 350
\bibitem{} Gerasimovi\u{c} B.P., 1933, Zeitschrift Astrophys., 7, 23
\bibitem{} Hamuy M. et al., 2003, Nature, 424, 651
\bibitem{} Hueckstaedt R.M., 2003, NewA, 8, 295
\bibitem{} H\"{o}flich P., Khokhlov A.M., 1996, ApJ, 457, 500
\bibitem{} Iben I. Jr., Renzini A., 1983, Annual Rev. Astron. Astrophys.
21, 271
\bibitem{} Itikawa Y., Ichimura A., 1990, J. Phys. Chem. Ref. Data, 19, 637
\bibitem{} Iwamoto K., Brachwitz F., Nomoto K., Kishimoto N., 
Umeda H., Hix W.R., Thielemann F.-K., 1999, ApJS, 125, 439
\bibitem{} Jeffery D.J., 1993, ApJ, 415, 734
\bibitem{} Khokhlov A.M., M\"{u}ller E., H\"{o}flich P., 1993, 
A\&A, 270, 223.
\bibitem{} Kupka F., Piskunov N., Ryabchikova T.A., Stempels H.C. \&
Weiss W.W., 1999, A\&AS, 138, 119
\bibitem{} Livio M., Riess A., 2003, ApJL, 594, L93
\bibitem{} Lundqvist P., Fransson C. 1988, A\&A, 192, 221
\bibitem{} Madej J., Nale\.{z}yty N., Althaus L.G., 2004, A\&A, 419, L5
\bibitem{} Prada Moroni P.G., Straniero O., 2003, Mem.S.A.It., 74, 508 
\bibitem{} Rigon L. et al., 2003, MNRAS, 340, 191
\bibitem{} Salim S., Rich R.M., Hansen B.M., Koopmans L.V.E., 
Oppenheimer B.R., Blandford R.D., 2004, ApJ, 6001, 1075
\bibitem{} Sreenivasan K.R., Ramshankar R., Meneveau C., 1989, Proc. R.
Soc. Lond. A 421, 79
\bibitem{} Stockdale J., Sramek R.A., Weiler, K.W., Van Dyk, S.D., Panagia, N., 
2003, IAUC No. 8157
\bibitem{} Stolarski R.S., Green A.E.S., 1967, JGR, 72, 3967
\bibitem{} Turatto M. et al., 2000, ApJL, 534, L57
\bibitem{} Wang L., Baade D., H\"{o}flich P., Wheeler J.C., Kawabata K.,
Nomoto K., 2004, ApJ, 604, L53
\bibitem{} Whelan J., Iben I. Jr., 1973, ApJ, 186, 1007
\bibitem{} Weiler K.W., Sramek R.A., Panagia N., van der Hulst J.M., 
Salvati M., 1986, ApJ, 301, 790
\bibitem{} Wood-Vasey, W.~M., et al., 2002, IAUC No. 8019, 2
\bibitem{} Zijlstra A.A., 2004, MNRAS, 348, L23


\end{thebibliography}
\end{document}